\documentclass[reprint,prb,aps,superscriptaddress,showpacs,twocolumn]{revtex4-1}
\usepackage{epsfig,color}
\usepackage{graphicx}
\usepackage{amsmath}
\usepackage{amssymb}

\begin{document}
\title{Spontaneous and resonant lifting of the spin blockade in nanowire quantum dots}

\author{M. P. Nowak and B. Szafran}
\address{AGH University of Science and Technology, Faculty of Physics and Applied Computer Science,\\
al. Mickiewicza 30, 30-059 Krak\'ow, Poland}

\begin{abstract}
A complete numerical description of the charge and spin dynamics of a two-electron system confined in narrow nanowire quantum dots under oscillating electric field is presented in the context of recent electric dipole spin resonance experiments. We find that the spin-orbit coupling results in lifting the spin blockade by phonon mediated relaxation provided that the initially occupied state is close in energy to the ground state. This leads to suppression of the blockade from the triplet state with spins polarized parallel to the external magnetic field $B$. At higher $B$, after singlet-triplet ground-state transition a new channel for lifting the Pauli blockade opens which results in an appearance of additional resonance lines. The calculated signatures of this transition are consistent with recent experimental results [S. M. Frolov et al., Phys. Rev. Lett. {\bf 109}, 236805 (2012)].
\end{abstract}


\maketitle
\section{Introduction}
Coherent spin control is one of the necessary prerequisites for fabrication of solid-state quantum computer operating on spin qubits. Recently gate defined nanowire \cite{sostrnanowire} double quantum dots have been successfully used for experimental demonstration \cite{nadj-perge,schroer,nadj2012,frolov,petersson,berg,pribiag} of electrical control of confined spins.\cite{nowack} The spin rotations are performed by means of electric dipole spin resonance (EDSR) where spin-orbit (SO) interaction \cite{rashba,dresselhaus} is used for electrical control of spin excluding the need for introducing an oscillating magnetic field in the device.\cite{koppens} The spin oscillations are probed with spin blockade \cite{ono} of a two-electron system where the current cycle $(0,1)\rightarrow(1,1)\rightarrow(0,2)\rightarrow(0,1)$ [the numbers denote number of electrons in the adjacent quantum dots] is blocked at the $(1,1)\rightarrow(0,2)$ transition when the spin configurations of the $(1,1)$ and $(0,2)$ states do not match. For low bias, the only available $(0,2)$ state is the spin singlet so the current is blocked if the system is initialized in one of the spin-polarized triplets. The blockade is lifted when the spin of driven electron is flipped with the total spin of the $(1,1)$ state changed from $S=1$ to $S=0$. Relaxation of the $(1,1)$ state to $(0,2)$ singlet that follows the spin rotation is a consequence and signature of Pauli blockade removal and is the main phenomenon studied in this work. The $(1,1)\rightarrow(0,2)$ transition requires dissipation of the excess energy which is absorbed by the crystal vibrations.

SO coupling besides allowing for electric control of the spin leads to spin relaxation \cite{golovach,stano,meunier,danon2}  which is mediated by phonons. In low magnetic fields the spin relaxation occurs via both the hyperfine field and the spin-orbit coupling. EDSR experiments performed in conditions of the Pauli blockade require application of an external magnetic field to induce the spin Zeeman splitting of electron energy levels. Since the Zeeman splitting of the nuclear levels is much smaller, the direct exchange of spins between the electron and the nuclei is suppressed at a fields of the order of mT.\cite{nadj2010,pfund,danon,nori}
The fluctuations of the nuclear field occurs at the timescales of 10-100 microsecond \cite{hanson} while in the present work we focus on the spin evolution in a time scale of order of tenths on nanoseconds. Still static hyperfine field\cite{inni,johnson} can be used to mediate EDSR transitions in GaAs quantum dots\cite{laird2007} where the SO interaction is weak and can manifest itself by the compensation of the g-factor difference between the dots.\cite{frolov} In the present work we account for the effect of a spatially varied magnetic field by introducing the spatial dependence of the $g$ factor.

The spin transitions in EDSR driven by SO coupling \cite{nowak2012a} or hyperfine field \cite{osika} were the subject of our previous studies.
In this paper we describe the mechanism of transitions for the phonon field actively contributing to the process of the Pauli blockade lifting.
We are not only interested by the driven spin transitions as in previous papers,\cite{nowak2012a,osika} but we also investigate their consequence, i.e.
passage of both electrons to the quantum dot with deeper confinement potential after energy dissipation by the phonon field, which triggers the current flow
after the Pauli blockade is lifted.
 We indicate several features that are crucial for the mechanism of the current blockade lifting: $i$) the spin non-conserving relaxation $(1,1)\rightarrow(0,2)$ from triplet\cite{comm} state with spins polarized along the magnetic field ($|\uparrow,\uparrow\rangle$) is of a very similar effectiveness as the spin conserving $(1,1)\rightarrow(0,2)$ relaxation. $ii$) for small magnetic fields where $(0,2)$ singlet is the ground state the spontaneous spin relaxation from the $|\uparrow,\uparrow\rangle$ state contributes significantly to the lifting of the Pauli blockade.  $iii$) on the other hand the relaxation from the triplet with spins oriented antiparallel to the magnetic field orientation ($|\downarrow,\downarrow\rangle$) is by two orders of magnitude slower so the blockade is maintained. $iv$) at higher magnetic fields when the ($|\uparrow,\uparrow\rangle$) triplet becomes the ground state the spin rotation accompanied by charge redistribution results in lifting the spin blockade through a direct transition to $(0,2)$ singlet. We indicate the footprint of the latter mechanism in a recent experimental map.\cite{frolov}

\section{Theory}
The considered two-electron system is described by the Hamiltonian $H=\sum_ih^i(t)+e^2/(4\pi\varepsilon\varepsilon_0|\textbf{r}_1-\textbf{r}_2|)$ where $h^i$ is single electron energy operator.
We assume that the electrons are strongly localized near the axis of the wire and that they occupy the ground state of lateral quantization.
 We assume that the lateral wave function have a Gaussian form $\psi(y,z)=(\sqrt{\pi}l)^{-1}\exp[-(y^2+z^2)/2l^2]$, with $l=20$ nm for which an analytical form of the electron-electron interaction \cite{bednarek}
 can be derived upon integration of $H$ over the directions perpendicular to the wire,
\begin{equation}
H_{1D}=h^1_{1D}+h^2_{1D}+\frac{\sqrt{\pi/2}}{4\pi\varepsilon_0\varepsilon l}\mathrm{erfcx}\left[\frac{|x_1-x_2|}{\sqrt{2}l}\right],
\label{h2e}
\end{equation}
with the single-electron energy operator
\begin{equation}
h_{1D} = \frac{\hbar^2 k_x^2}{2m^*} + V(x) - \alpha \sigma_y k_x + \frac{1}{2}\mu_B g(x)B\sigma_x,
\label{1dham}
\end{equation}
where $\hbar k_x=-i\hbar\nabla_x$ is momentum operator and $H_{SO 1D}=-\alpha\sigma_yk_x$ stands for Rashba SO coupling which results from averaging the $H_{SO}=\alpha(\sigma_xk_y-\sigma_yk_x)$ Hamiltonian in the $y$-direction. We consider the nanowire grown in $[001]$ crystal direction.

We allow for position dependent $g$-factor in the device \cite{nadj-perge,nadj2012,frolov} and take $g(x)=g[1+\beta H(x)]$ where $H(x)$ is Heavyside step function and $\beta=0.1$, i.e. $g$-factor in the right dot is 1.1 of the value in the left dot. $V(x)=V_{c}(x)+eF_{bias}x$ is the stationary potential, where  $V_{c}$ defines potential of two quantum dots of $138$ nm width separated by potential barrier of $25$ nm width and $40$ meV height,
and $F_{bias}$ is the electric field setting the energy difference between the dots.
 The EDSR is induced by an oscillating electric field which is introduced by an
extra potential $V'(x,t)$. The driving AC electric field is assumed present in the left dot,\cite{nadj-perge} so the time dependent part of the potential takes the form $V'(x,t)=eF_{AC}xf(x)\sin(\omega_{AC}t)$ where $f(x)=1$ in the left dot and 0 outside -- see the inset to Fig. \ref{fig1}(a).

As the initial states for the time evolution we set one of the eigenstates of operator  (1).
The eigenstates of the Hamiltonian (\ref{h2e}) are determined with the configuration interaction scheme. In the applied approach the $n$-th two-electron spin-orbital is constructed in a basis consisting slater determinants, i.e.,
\begin{eqnarray}
\Psi^n(x_1,\sigma_1,x_2,\sigma_2,t=0)=\nonumber\\
\sum_i^\Lambda\sum_{j=i+1}^\Lambda A_{ij}^n\left[\psi_i(x_1,\sigma_1)\psi_j(x_2,\sigma_2)\nonumber\right.\\\left.-\psi_i(x_2,\sigma_2)\psi_j(x_1,\sigma_1)\right]
\end{eqnarray}
where the coefficients $A^n_{ij}$ are found by diagonalization of Hamiltonian (\ref{h2e}). Spin-orbitals $\psi(x,\sigma)$ are found by exact diagonalization of $h_{1D}$ on a mesh with $2\times201$ points.

For the description EDSR we solve the two-electron Schr\"odinger equation for Hamiltonian
\begin{equation}
H_{1D}(t)=H_{1D}+H'_{1D}(t),
\end{equation}
where $H_{1D}$ is the time independent part given by Eq. (1) and $H'_{1D}(t)=eF_{AC}[x_1f(x_1)+x_2f(x_2)]\sin(\omega_{AC}t)$ contains the oscillating electric field $F_{AC}$.
The time evolution is described in the basis of Eq. (1) eigenstates $\Psi^n(x_1,\sigma_1,x_2,\sigma_2)$ corresponding to eigenenergies $E^n$.
The two-electron spinor is expressed as
\begin{equation}
\begin{split}
\Psi(x_1,\sigma_1,x_2,&\sigma_2,t)=\\&\sum_n^N c_n(t)\exp(-iE_nt/\hbar)\Psi^n(x_1,\sigma_1,x_2,\sigma_2).
\end{split}
\label{rr}
\end{equation}
 Eq. (\ref{rr}) plugged into the Schr\"odinger equation gives a system of linear differential equations
 for time evolution of coefficients $c_n$
\begin{equation}
\frac{d}{dt}c_n(t)=-\frac{i}{\hbar}\sum_{m=1}^N c_m(t)\langle\Psi^n|H_{1D}'(t)|\Psi^m\rangle.
\end{equation}
We use $N=20$ basis states in Eq. (\ref{rr}) which provides numerically accurate results as compared to a direct finite difference solution of the time dependent Schr\"odinger equation in the above system.\cite{nowak2012a}

The form of the wave function (\ref{rr}) developed in the basis is convenient for simulation of the energy dissipation by phonons.
In our modeling we include the transitions between the two-electron states due to bulk phonon mediated relaxation with a rate given by the Fermi golden rule. The relaxation rate between the initial $\Psi^\imath$ and final $\Psi^f$ states is described by,
\begin{equation}
\begin{split}
\tau_{\imath f}^{-1}&=\frac{2\pi}{\hbar}\sum_{\nu,\xi=1,2}\int_{\mathbf{q}}dq|M_\nu(\mathbf{q})|^2\\&\times|\langle\Psi^f|e^{-i\mathbf{qr_i}}|\Psi^\imath\rangle|^2\delta(|E^f-E^\imath|-E_q),
\label{fermi}
\end{split}
\end{equation}
where the phonon dispersion relation is $E_q=\hbar c_{\nu}|\mathbf{q}|$ and $c_{\nu}$ is the sound velocity. The sum in (\ref{fermi}) goes over three types of electron-phonon scattering ($\nu$) due to: deformation potential with longitudinal mode \cite{bockelmann} ($\nu=$LA-DP) with,
\begin{equation}
|M_{LA-DP}(\mathbf{q})|^2=\frac{\hbar D^2}{2dc_{LA}}|\mathbf{q}|,
\end{equation}
where $D$ stands for the crystal acoustic deformation potential constant, $d$ is mass density, and $c_{LA}$ is sound velocity of phonon LA mode.
Electron-LA phonon scattering due to the piezoelectric field \cite{climente2006} ($\nu=$LA-PZ),
\begin{equation}
|M_{LA-PZ}(\mathbf{q})|^2=\frac{32\pi^2\hbar e^2 h_{14}^2}{\varepsilon^2dc_{LA}}\frac{(3q_xq_yq_z)^2}{|\mathbf{q}|^7},
\end{equation}
where $h_{14}$ is PZ constant and electron-TA phonon scattering due to the piezoelectric field ($\nu=$TA-PZ),\cite{climente2006}
\begin{equation}
\begin{split}
|M_{TA-PZ}(\mathbf{q})|^2&=2\times\frac{32\pi^2\hbar e^2h_{14}^2}{\varepsilon^2dc_{TA}}\left|\frac{q_x^2q_y^2+q_y^2q_z^2+q_z^2q_x^2}{|\mathbf{q}|^5}\right.\\&\left.-\frac{(3q_xq_yq_z)^2}{|\mathbf{q}|^7}\right|,
\end{split}
\end{equation}
where the multiplication by two results from two transverse phonon modes.

The relaxation is included into time dependent calculation in the following way. The $|c_n(t)|^2$ values are changed in each time step with an account taken for relaxation to all lower energy states and from all the higher energy states that span the basis. The relaxation is simulated using the following formula
\begin{equation}
|c_n(t+\Delta t)|^2=|c_n(t)|^2+\sum_{m=n+1}^N\tau^{-1}_{mn}|c_m|^2\Delta t-\sum_{m=1}^{n+1}\tau^{-1}_{nm}|c_n|^2 \Delta t.
\label{fonevo}
\end{equation}
The formula allows for relaxation and not absorption of the energy by the electrons which is equivalent to assumption that the system is kept in a 0K bath.

Time dependent calculations are performed taking on the same footing evolution due to Eq. (\ref{fermi}) and Eq. (\ref{fonevo}), i.e. in each step of the time evolution the coefficients $c_n$ are changed due to oscillating electric field and phonon-mediated relaxation.

We assume material parameters for InSb, i.e. electron effective mass $m^*=0.014$, $g=-51$, dielectric constant $\varepsilon = 16.5$ and take the Rashba constant $\alpha=10$ meVnm. The AC field amplitude $F_{AC} = 0.05\; \mathrm{kV/cm}$ is assumed. For calculation of phonon mediated relaxation we take \cite{defpot} $D=5775\;\mathrm{kg}/\mathrm{m}^3$, $h_{14}=1.41\times10^9$ V/m after [\onlinecite{climente2006}] and we take sound velocities: $c_{LA}=3.8\times10^3$ m/s after [\onlinecite{hebboul}] and  $c_{TA}=1.9\times10^3$ m/s from Ref. \onlinecite{landolt}. We use basis consisting of $\Lambda=50$ single-electron orbitals which provides accuracy of two-electron energy levels better than $0.5\;\mu$eV for $B=0.11$ T.

\section{Results}

\begin{figure}[ht!]
\epsfxsize=80mm
                \epsfbox[34 160 550 690] {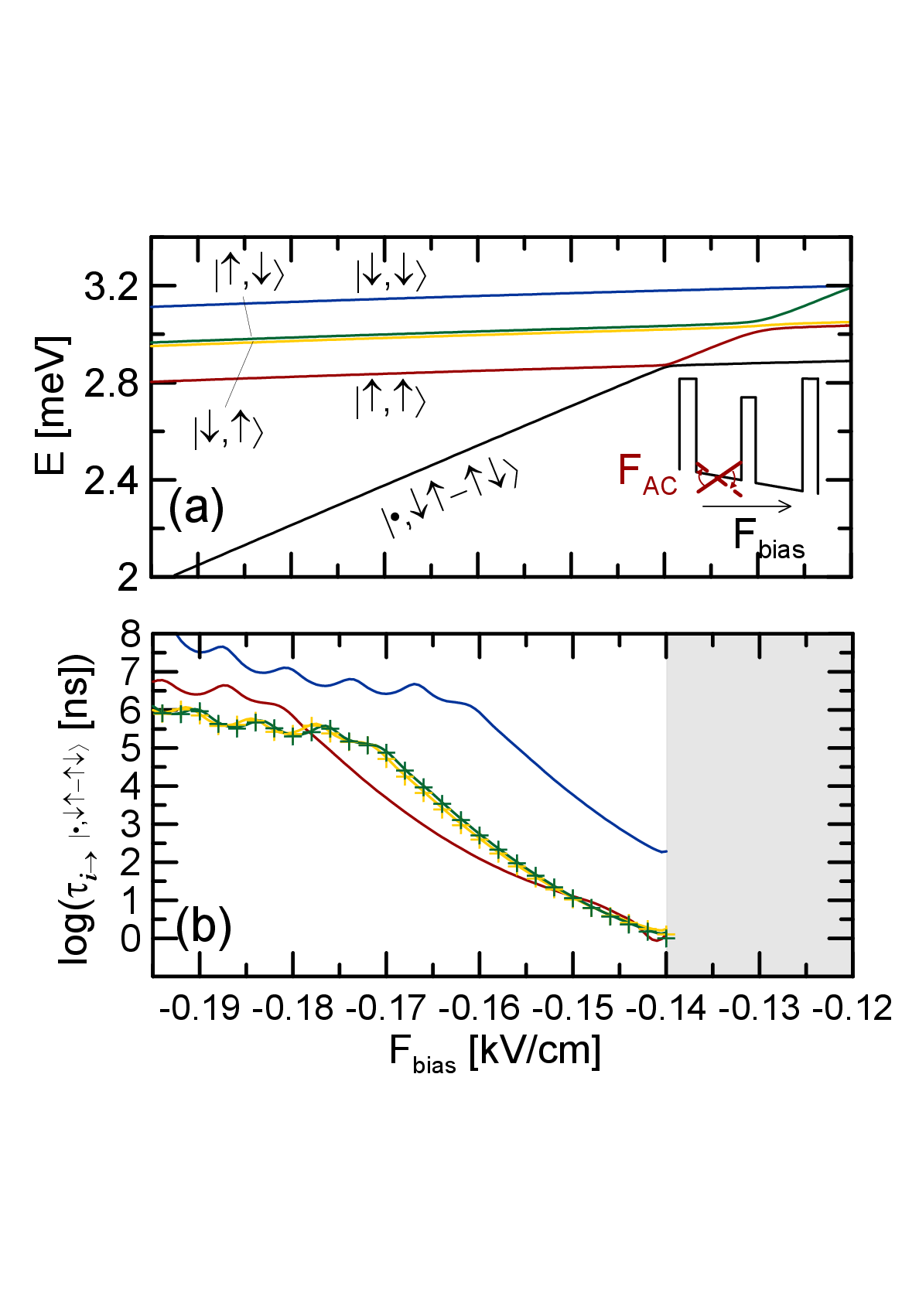}
                 \caption{(a) Energy levels of two-electron double quantum dot as a function of bias electric field for $B=50$ mT in the presence of spin-orbit interaction. The arrows illustrate approximate spin polarization of electrons in the dots. The inset present schematics of the considered confinement potential. (b) relaxation time of excited states due to phonon mediated relaxation to the (0,2) singlet state. The colors of curves denote the initial state of relaxation. The symbols (curves) corresponds to the results obtained without (with) SO interaction.}
 \label{fig1}
\end{figure}

The charge distribution in the double dot is controlled by external voltages applied along the structure. Fig. \ref{fig1}(a) presents the lowest part of the energy spectrum obtained in the presence of SO coupling (the subsequent energy levels -- of (0,2) triplets -- are above 5 meV) as a function of bias electric field for $B=50$ mT. For the most negative values of $F_{bias}$ the ground state is a singlet $|\bullet,\downarrow\uparrow-\uparrow\downarrow\rangle$ in which both electrons reside in the right dot [(0,2) configuration]. With the ($\bullet$) in the bracket we mark the unoccupied left dot. The four excited states correspond to single occupancy of each dot [(1,1) configuration] and energy of those states only weakly change as a function of bias electric field. The two states close in energy, i.e. $|\downarrow,\uparrow\rangle$ and $|\uparrow,\downarrow\rangle$ correspond to definite and opposite spin configurations in each dot (i.e., in the $|\downarrow,\uparrow\rangle$ state spin of the electron in the right dot -- where the $g$-factor takes the highest value -- is polarized along the magnetic field) resulting from mixing of the spin-zero triplet with the singlet state by the $g$-factor mismatch between the dots and negligible exchange coupling. The two triplets $|\uparrow,\uparrow\rangle$, $|\downarrow,\downarrow\rangle$ are split by Zeeman interaction.

\begin{figure}[ht!]
\epsfxsize=75mm
                \epsfbox[22 270 570 570] {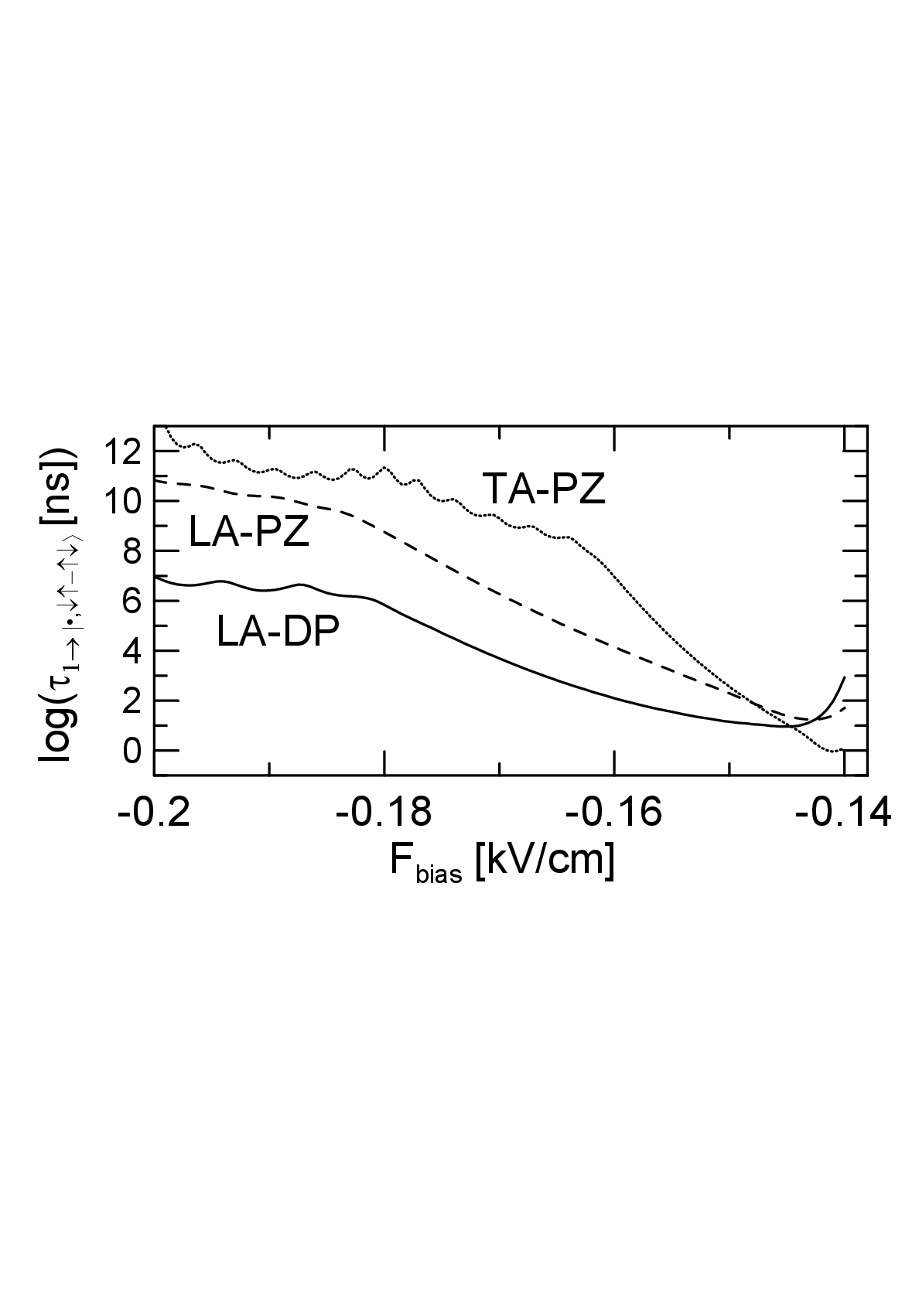}
                 \caption{Relaxation time of the $|\uparrow,\uparrow\rangle$ to ground state singlet mediated by different electron-phonon scattering types.}
 \label{fig2}
\end{figure}

In Fig. \ref{fig1}(b) we present relaxation times $\tau_{i\rightarrow|\bullet,\downarrow\uparrow-\uparrow\downarrow\rangle}$ of excited states to the ground state singlet. In the absence of SO interaction phonon scattering couples only states with the same total spin. In that case only the relaxation times of $|\uparrow,\downarrow\rangle$ and  $|\downarrow,\uparrow\rangle$ have finite values and they are presented with the crosses in Fig. \ref{fig1}(b). At low values of $F_{bias}$ the relaxation times are of order of milliseconds but when the energy differences between the initial states and (0,2) singlet become lower the relaxation times rapidly drop allowing for $(1,1)\rightarrow(0,2)$ spin-conserving relaxation within nanoseconds for $F_{bias}>-0.16$ kV/cm.

\begin{figure*}[ht!]
\epsfxsize=180mm
                \epsfbox[26 326 570 516] {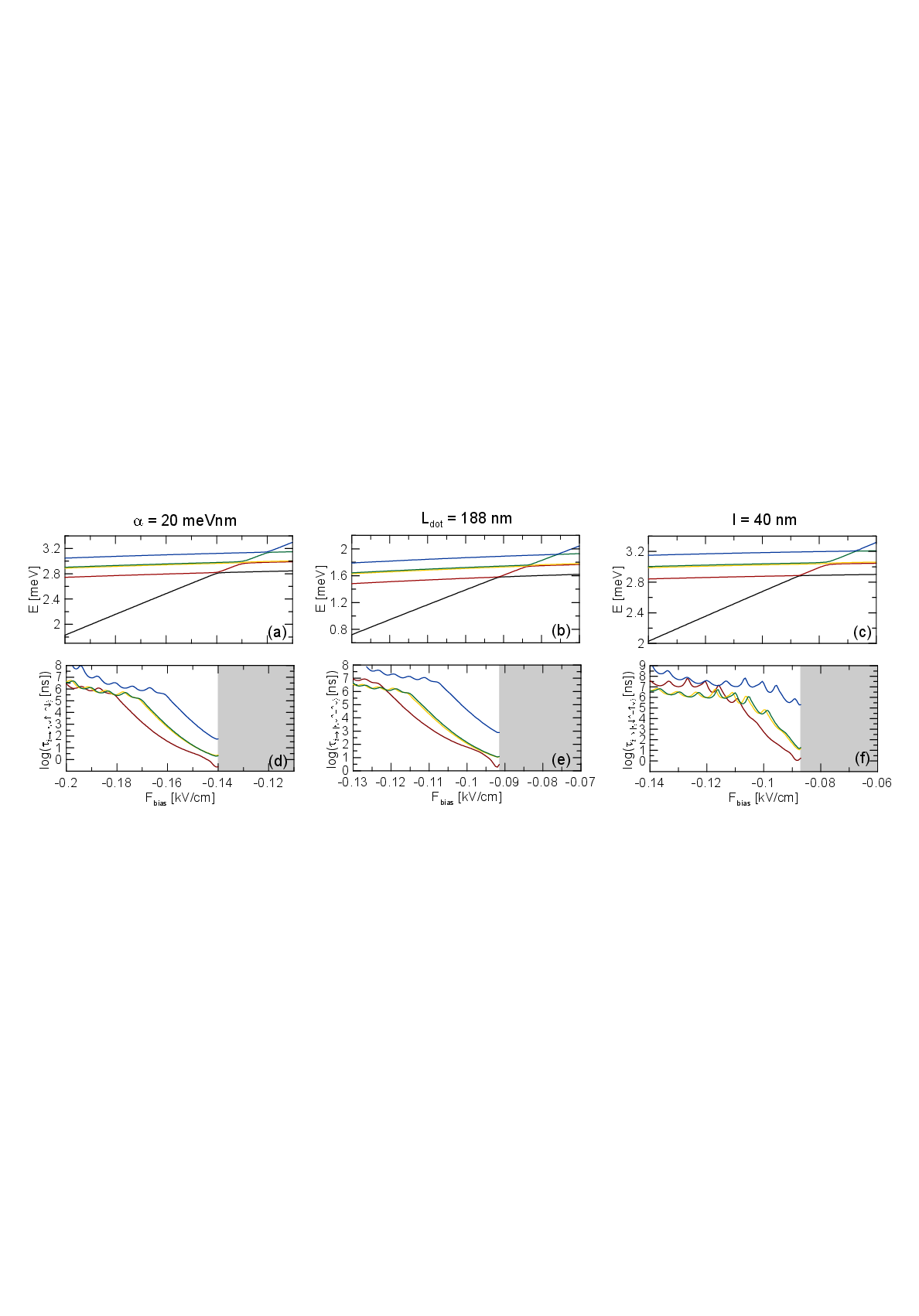}
                 \caption{Lowest part of the energy spectrum obtained for (a) spin-orbit interaction constant increased to $\alpha=20$ meVnm (b) length of each dot increased by 50 nm and (c) increased radius of the nanowire that results in an spread of wave function in the lateral direction such $l=40$ nm. (d)-(f) present relaxation times of excited states to $|\bullet,\downarrow\uparrow-\uparrow\downarrow\rangle$ obtained for parameters corresponding to the upper plots.}
 \label{fig3}
\end{figure*}

When SO coupling is included the spin polarization of the states becomes only approximate. The relaxation times of $|\downarrow,\uparrow\rangle$ and $|\uparrow,\downarrow\rangle$ states do not change -- see the curves and crosses in Fig. \ref{fig1}(b). However now relaxation from all the (1,1) states to (0,2) singlet is open. For most negative values of $F_{bias}$ relaxation time of triplet states is longer than tenths of milliseconds but when the bias field is increased the relaxation time of $|\uparrow,\uparrow\rangle$ triplet becomes about the same as the two spin opposite states. Relaxation time of $|\downarrow\downarrow\rangle$ state are longer by two orders of magnitude from the rest of the (1,1) states.

For low values of $F_{bias}$ where the relaxation times are of order of milliseconds and more one can observe ripples in the curves. The bias difference between two ripples corresponds to a change in the energy of $\Delta E\simeq 0.106$ meV. The latter corresponds to an increase of the wavelength of LA-DP phonon -- the change in wavelength of the oscillatory part in matrix element in Eq. (\ref{fermi}) -- by $\sim150$ nm which is half of the length of double dot. Therefore the ripples are connected with the presence of an integer phonon wave within a single quantum dot.

In Fig. \ref{fig2} we compare the impact of individual electron-phonon coupling types on the relaxation time of the $|\uparrow,\uparrow\rangle$ to the $|\bullet,\downarrow\uparrow-\uparrow\downarrow\rangle$ state. We observe that LA-DP scattering dominates for almost all values of $F_{bias}$. Only when the energy separation between $|\uparrow,\uparrow\rangle$ triplet and the $|\bullet,\downarrow\uparrow-\uparrow\downarrow\rangle$ singlet becomes small [see Fig. \ref{fig1}(a)] the TA-PZ term starts to dominate giving relaxation times $\tau_{1\rightarrow|\bullet,\downarrow\uparrow-\uparrow\downarrow\rangle}\simeq1$ ns.

Discussed short relaxation times of the $|\uparrow\uparrow\rangle$ state are not specific to a particular parameter set as we checked for different strengths of SO coupling and the dot size. We calculated relaxation times of excited states as a function of the bias voltage for different value of SO coupling strength [see Figs. \ref{fig3}(a),(d)], length of the dots [see Figures \ref{fig3}(b),(e)] and the nanowire radius that controls the spread of lateral Gaussian wave function [see Figs. \ref{fig3}(c),(f)]. We observe that in each case the situation is generally the same as described previously, i.e., either the relaxation from all excited states is of the order of milliseconds at least or the relaxation from $|\uparrow,\uparrow\rangle$ triplet state with spins polarized along the magnetic field is faster than relaxation from $|\downarrow,\uparrow\rangle$ and $|\uparrow,\downarrow\rangle$ states.


\begin{figure}[ht!]
\epsfxsize=80mm
                \epsfbox[34 130 560 730] {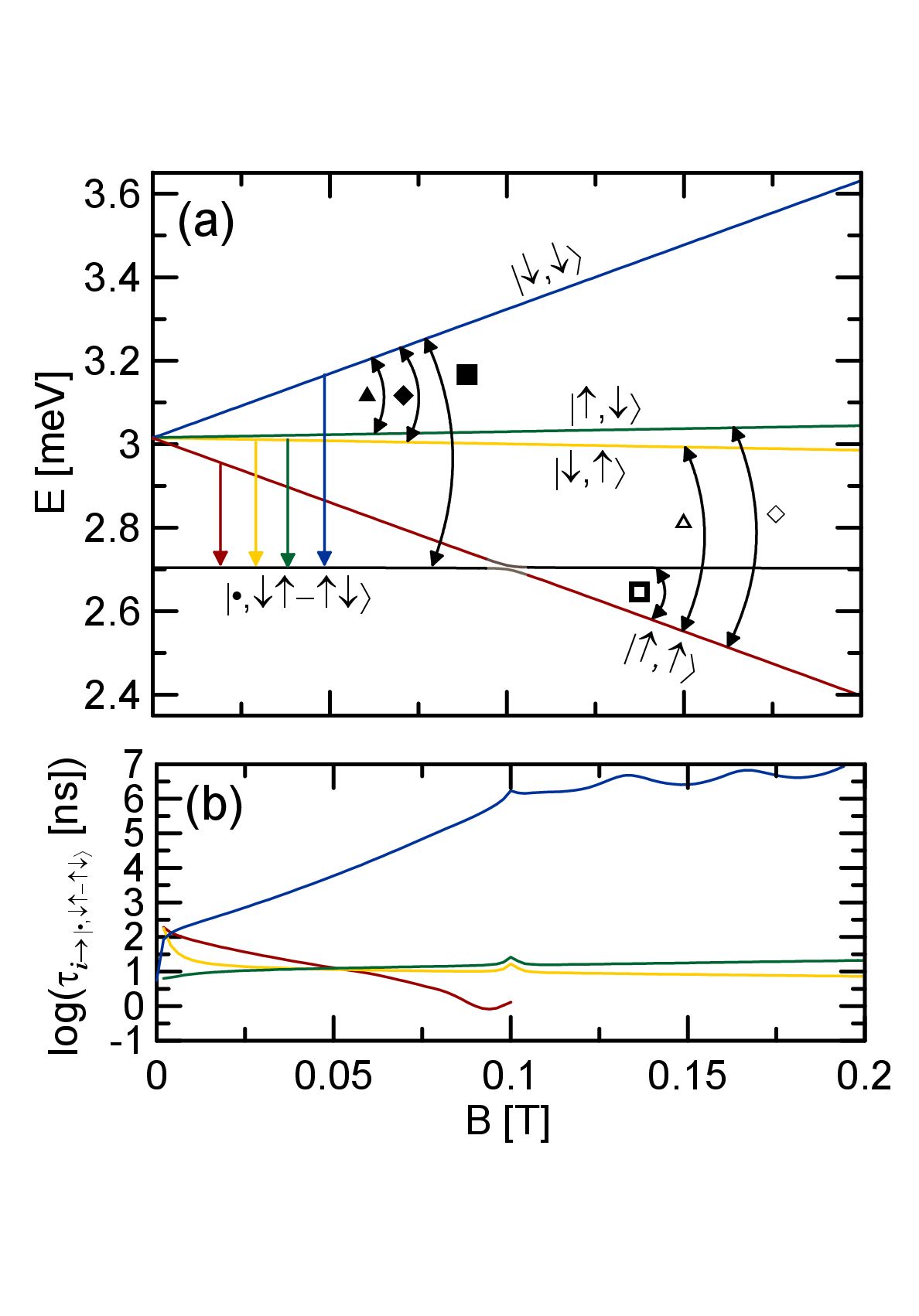}
                 \caption{(a) Energy spectrum as a function of the magnetic field. Straight arrows denote transitions due to phonon mediated relaxation. Curved arrows depict available EDSR resonances from the triplets. (b) Relaxation times of excited states. Results obtained for $F_{bias}=-0.15$ kV/cm.}
 \label{fig4}
\end{figure}

\begin{figure*}[ht!]
\epsfxsize=180mm
                \epsfbox[20 330 580 515] {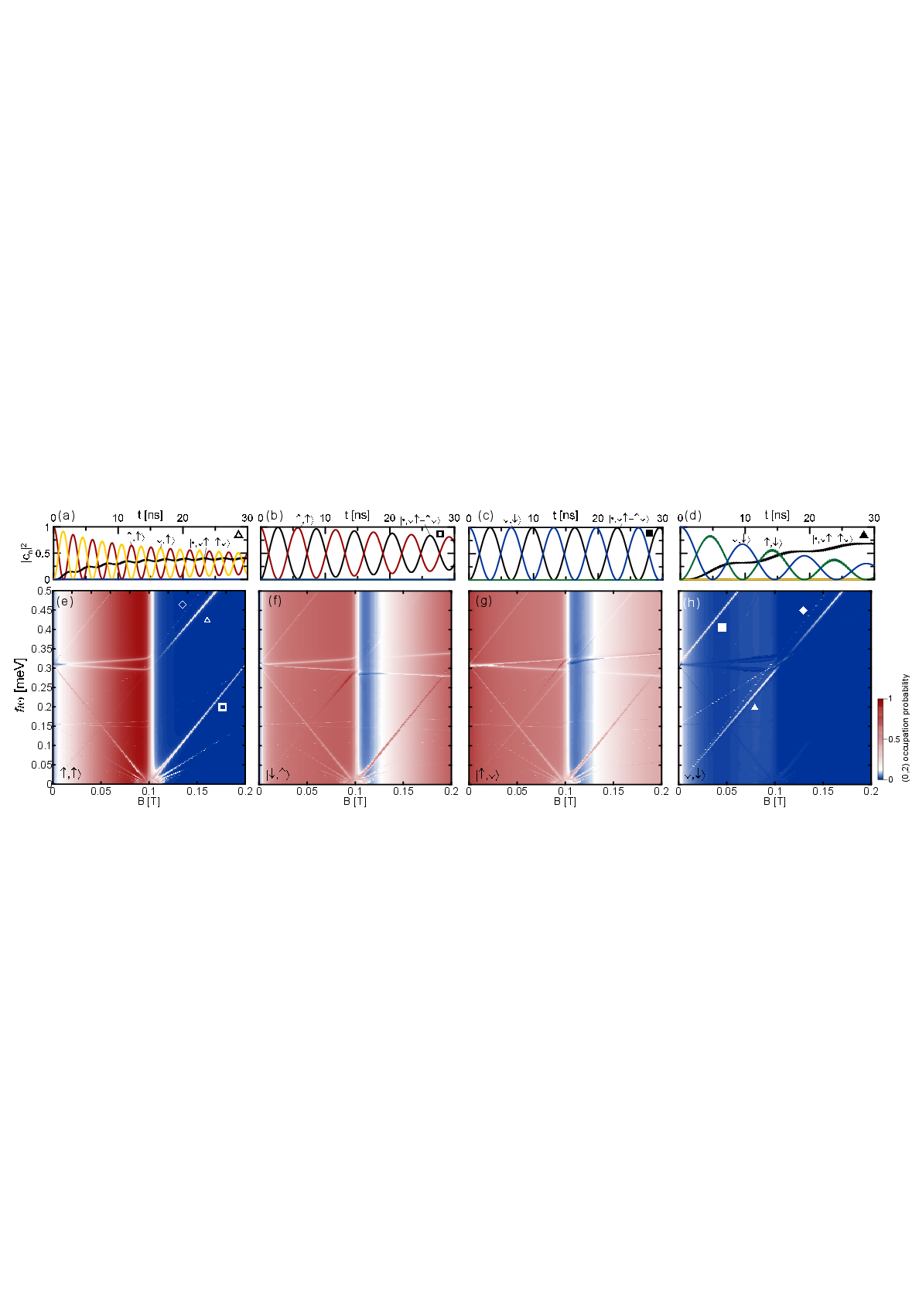}
                 \caption{(a)-(d) Transitions between the eigenstates during the time evolution at the resonances marked with the symbols. (e)-(h)  Probability of (0,2) occupation averaged during the 30 ns time evolution obtained for subsequent (1,1) states taken as the initial state of the time evolution -- the spin configuration of the initial state is denoted in the left bottom corner of each plot.}
 \label{fig5}
\end{figure*}
The experimental studies \cite{berg} report spin coherence time of order of tenths nanoseconds and coherent manipulation over a single spin up to 100 ns. We therefore focus on $F_{bias}$ range where the relaxation times are of order of tenths of nanoseconds -- hereafter we take $F_{bias}=-0.15$ kV/cm -- that allow for deblocking of single-electron current through the double dot after spin rotation in EDSR experiments. In Fig. \ref{fig4}(a) we plot energy levels as functions of the magnetic field. For $B=0$ the excited state is fourfold degenerate due to high interdot barrier -- negligible exchange coupling. When the magnetic field is increased the energy levels of the two spin polarized triplets $|\uparrow,\uparrow\rangle$ and $|\downarrow,\downarrow\rangle$ are split by the Zeeman interaction. On the other hand the energy levels of the two spin-opposite states -- $|\downarrow,\uparrow\rangle$ and $|\uparrow,\downarrow\rangle$ -- are weakly split due to $g$-factor mismatch in the dots. At $B=0.1$ T an anticrossing between the energy levels of triplet $|\uparrow,\uparrow\rangle$ and $(0,2)$ singlet states appear followed by the change of the ground state.

Figure \ref{fig4}(b) presents relaxation times of excited states to (0,2) singlet. We observe that the relaxations from spin antiparallel  $|\downarrow,\uparrow\rangle$ and $|\uparrow,\downarrow\rangle$ states occur within few nanoseconds regardless of $B$ value. As the energy separation between energy levels of $|\uparrow,\uparrow\rangle$ and (0,2) singlet decreases the relaxation time drops and after $B=50$ mT the relaxation time of this state is even lower than the relaxation time of spin-antiparallel states. On the other hand the relaxation from  $|\downarrow,\downarrow\rangle$ state is slow and the relaxation time grows for increasing magnetic field until $B=0.1$ T.

In EDSR experiments the two-electron system can be initialized in any of the low energy states within the transport energy window. We therefore study the time evolution taking each of the (1,1) states with the energy levels depicted in Fig. \ref{fig4}(a) as the initial state. In Figs. \ref{fig5}(e)-(h) we present (0,2) occupation probability (which would allow for tunneling of one of the electrons outside the dot lifting the blockade) averaged during 30 ns time evolution as a function of the magnetic field and the electric field frequency $\omega$.

For $|\uparrow\uparrow\rangle$ taken as the initial state the averaged (0,2) occupation probability is presented in Fig. \ref{fig5}(e). At the left part of the map we observe increasing probability in the background as a function of $B$ due to spin relaxation that results in spontaneous lifting of spin blockade. At $B=0.1$ T the phonon mediated SO relaxation to singlet (0,2) from $|\uparrow,\uparrow\rangle$ stops as the latter becomes the ground state -- the background of the plot shows nearly zero (0,2) occupation probability. However we observe several resonance lines with an increased probability. The resonance line ($\vartriangle$) corresponds to the spin rotation in the left dot ($|\uparrow,\uparrow\rangle\rightarrow|\downarrow,\uparrow\rangle$) accompanied by the phonon mediated relaxation to the $|\bullet, \downarrow\uparrow-\uparrow\downarrow\rangle$ singlet which results in an increase of the $(0,2)$ occupation probability -- see Fig. \ref{fig5}(a) where we present the probability $|c_n|^2$ of finding the system in the $n$'th state during the time evolution. The ($\lozenge$) transition is related to the spin rotation in the right dot which is much less effective due to presence of the AC electric field only in the left dot and high interdot barrier which results in a narrow resonance line. The bottom line marked with ($\square$) corresponds to the direct transition to the $(0,2)$ singlet that involves charge reconfiguration between the dots -- see Fig. \ref{fig5}(b). Note that line of increased probability due to $|\uparrow,\uparrow\rangle\rightarrow|\bullet,\downarrow\uparrow-\uparrow\downarrow\rangle$ transition is not observed for $B<0.1$ T as in this region the spin relaxation of the triplet results in its fast deexcitation to the ground state with rate that exceeds the EDSR transition.

In Figures \ref{fig5}(f,g) one finds nonzero (0,2) occupation probabilities due to fast spin-conserving  relaxation of $|\downarrow,\uparrow\rangle$ and $|\uparrow,\downarrow\rangle$ to the $|\bullet,\downarrow\uparrow-\uparrow\downarrow\rangle$ state as discussed previously. In fact this relaxation is fast enough that one can observe lines of lowered probability when the system already relaxed into singlet (0,2) is driven back to one of the excited states.

For the $|\downarrow,\downarrow\rangle$ triplet taken as the initial state outside the resonances the (0,2) occupation probability is nearly zero at Fig. \ref{fig5}(h) as the phonon mediated relaxation from this state is slow -- see the blue curve in Fig. \ref{fig4}(b). This shows that for magnetic field range before the anticrossing only the $|\downarrow,\downarrow\rangle$ triplet provides spin blockade as the $|\uparrow,\uparrow\rangle$, $|\downarrow,\uparrow\rangle$ and $|\uparrow,\downarrow\rangle$ states decay quickly into $|\bullet,\downarrow\uparrow-\uparrow\downarrow\rangle$. The lines that go through the diagonal of the plot -- $(\blacklozenge)$, $(\blacktriangle)$ -- corresponds to the transition to the $|\downarrow,\uparrow\rangle$ and $|\uparrow,\downarrow\rangle$ states respectively accompanied by relaxation to $|\bullet,\downarrow\uparrow-\uparrow\downarrow\rangle$ [see Fig. \ref{fig5}(d)] and the line at the left upper part of the plot -- ($\blacksquare$) -- is a direct transition to the $(0,2)$ singlet that does not involve phonon mediated relaxation [see Fig.\ref{fig5}(c)].

Note that in maps of Figs. \ref{fig5}(a) and (d) also lines of increased probability at the half frequency of the ($\square$) and ($\blacksquare$) transitions are visible which is due to resonant harmonic generation by the driven electrons.\cite{nowak2012a}

\begin{figure}[ht!]
\epsfysize=75mm
                \epsfbox[30 170 555 661] {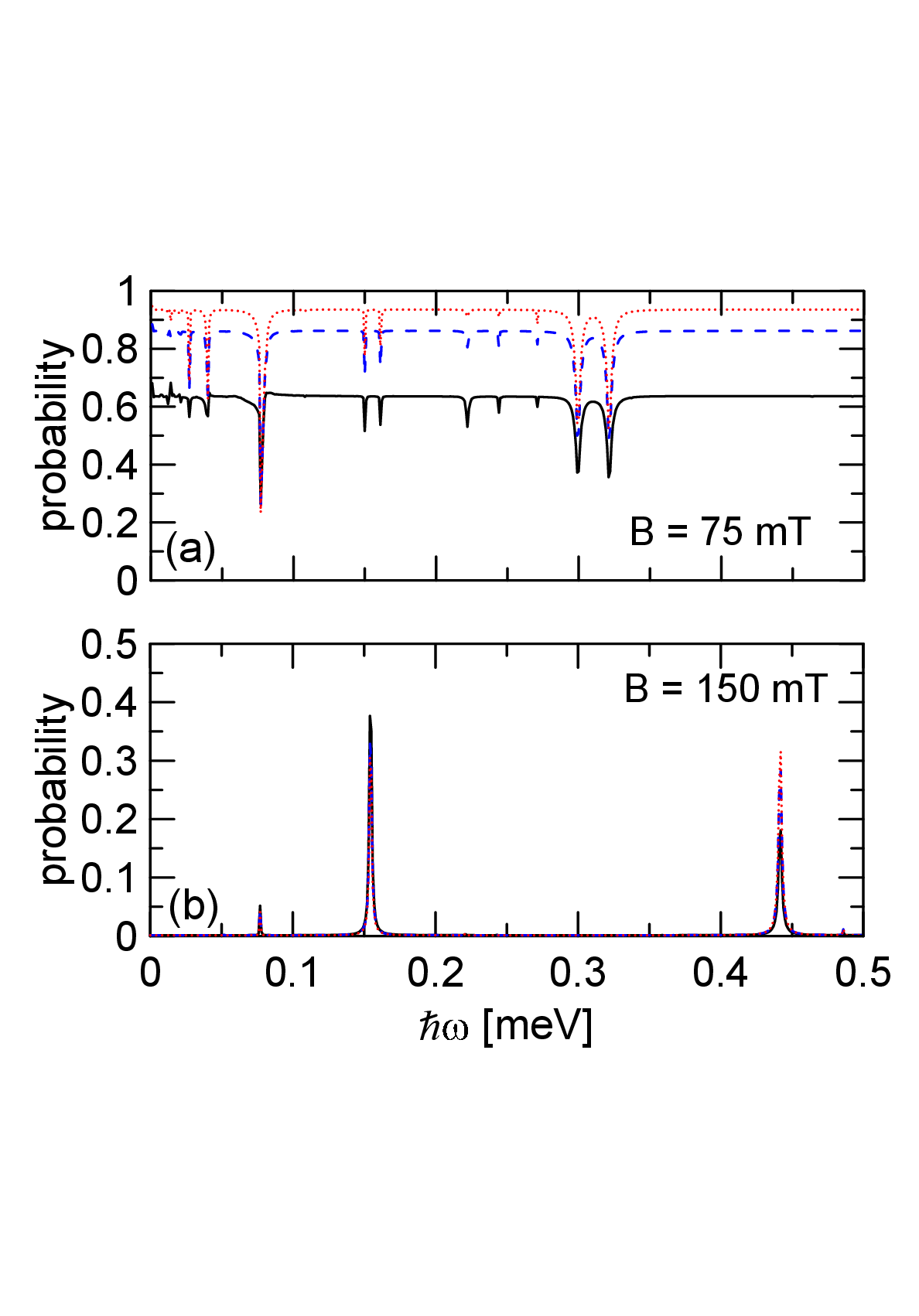}
                 \caption{Probability of (0,2) occupation obtained during 10 ns (black solid curves), 30 ns (blue dashed curves) and 70 ns (red dotted curves) time evolution before (a) and after (b) singlet-triplet anticrossing. The initial state is $|\uparrow,\uparrow\rangle$ triplet.}
 \label{fig6}
\end{figure}

In this work we set the duration of the time evolution to 30 ns.
The significance of the duration is presented in Fig. \ref{fig6}(a, b). For $B=75$ mT where the ground state is $|\bullet, \downarrow\uparrow-\uparrow\downarrow\rangle$ singlet the increase of the time evolution results in an increase of the (0,2) occupation probability as the system outside the resonance has enough time to relaxe to the ground state [compare the black and red curve in Fig. \ref{fig6}(a)]. On the other hand the peaks that corresponds to the lowered (0,2) occupation probability that appear due to the excitation to the higher energy levels are mainly not affected by the change of the evolution time as they corresponds to a constant excitation-relaxation process. For the magnetic field $B = 150$ mT where the $|\uparrow,\uparrow\rangle$ triplet is the ground state outside the resonances the (0,2) occupation probability is zero. The peak that corresponds to the excitation to the $|\downarrow,\uparrow\rangle$ becomes higher as the increased evolution time allows now for complete relaxation to $|\bullet, \downarrow\uparrow-\uparrow\downarrow\rangle$ [compare with Fig. \ref{fig5}(d) plotted for 30 ns].

\begin{figure}[ht!]
\epsfxsize=85mm
                \epsfbox[20 250 580 590] {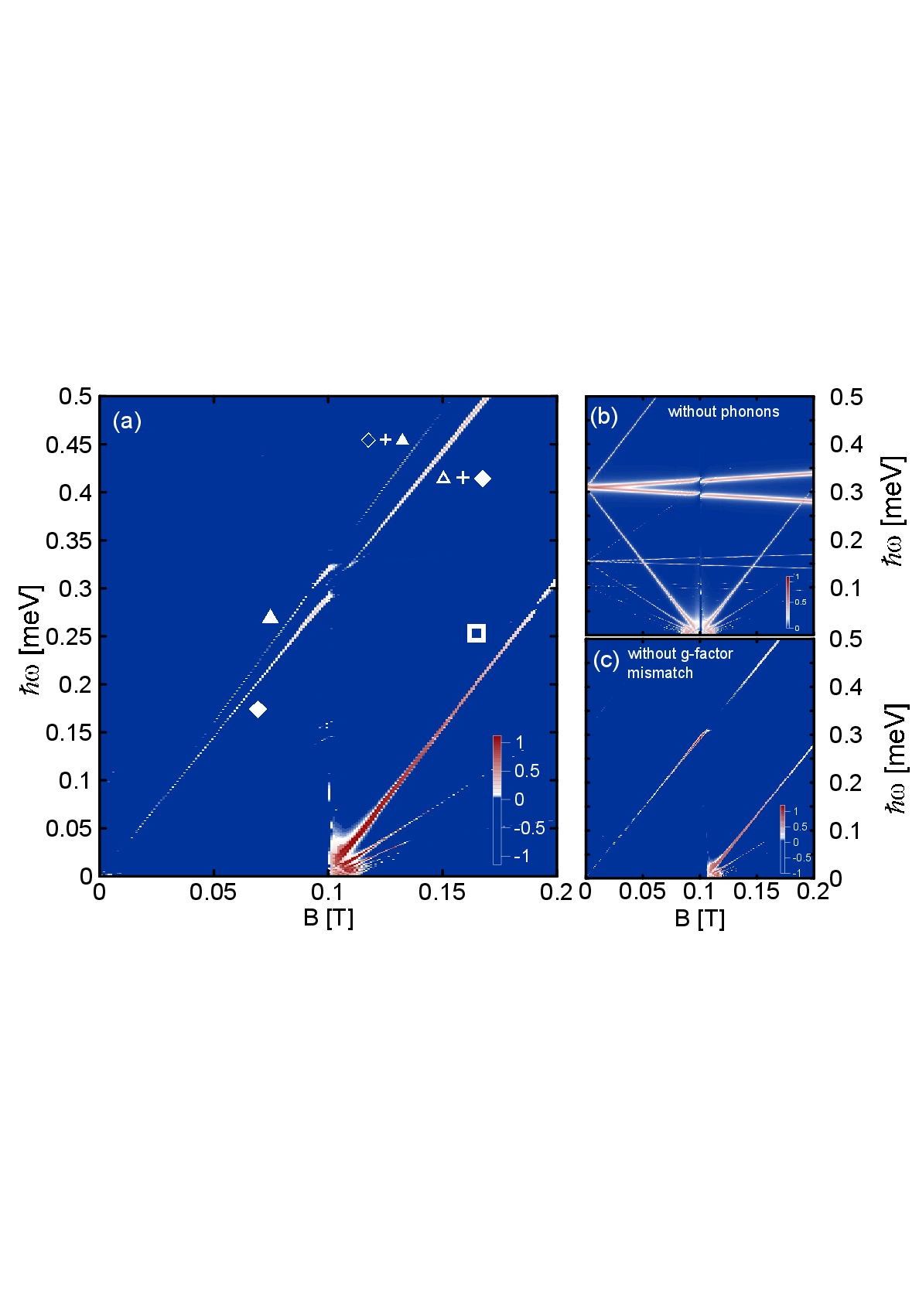}
                 \caption{(a) Probability of (0,2) occupation averaged over the 30 ns time evolution calculated as a sum of results obtained for initial states with (1,1) occupation. For each $B$ value the (0,2) occupation probability obtained for $\hbar\omega_{AC}=0$ (in the absence of driving electric field) was subtracted from the results. (b) Same as (a) but without phonon mediated relaxation. (c) Same as (a) but without g-factor difference between the dots.}
 \label{fig7}
\end{figure}

The experimentally  measured current maps\cite{frolov} are obtained from many sequential events of single electron transport through the structure. In each of them the system can initialize in any of the spin (1,1) states. We therefore calculate the total probability of (0,2) occupation by averaging the results over the initial states presented in Figs. \ref{fig5}(e)-(h) for 30 ns simulation time. For each value of $B$ the probability obtained without the oscillating electric field (due to pure relaxation) is subtracted to mimic the experimental procedure of Ref. \onlinecite{frolov} of removing the leakage current from the signal induced by AC field. The (0,2) occupation probability is displayed in Fig. \ref{fig7}(a). For low values of $B$ we observe two lines at the diagonal of the map that corresponds to the transitions from $|\uparrow,\uparrow\rangle$ state -- rotation of the spin down to spin up in the left dot (bright line) or in the right dot (faint line) accompanied by relaxation to (0,2) singlet. After singlet-triplet anticrossing at $B=0.1$ T the lines correspond to transition from {\it both} the triplets. At $B=0.1$ T additional resonance line starts at the bottom of the plot that corresponds to spin rotation with charge reconfiguration from $|\uparrow,\uparrow\rangle$ triplet. Note that there is no similar line corresponding to transition from the $|\downarrow,\downarrow\rangle$ state as it is compensated by the lowered probability obtained for evolution starting from $|\downarrow,\uparrow\rangle$ and $|\uparrow,\downarrow\rangle$.
In order to illustrate the impact of the phonon mediated relaxation on the lifting of the blockade in EDSR we calculated map of (0,2) occupation probability with neglected phonon mediated relaxation and display the results in Fig. \ref{fig7}(b). Now all of the resonances correspond to direct transition to $|\bullet,\downarrow\uparrow-\uparrow\downarrow\rangle$ induced by the AC electric field. We observe resonance lines which previously [compare with Fig. \ref{fig7}(a)] were masked by the {\it spontaneous} transition to the $|\bullet,\downarrow\uparrow-\uparrow\downarrow\rangle$ state and are not present in Fig. \ref{fig7}(a). Such lines are not present in the experimental maps.\cite{nadj-perge,schroer,nadj2012,frolov,petersson,berg,pribiag} Moreover the resonance lines at the diagonal of the plot which are found in all the experimental maps are present exclusively for active phonon mediated relaxation as it allows for the decay of the $|\uparrow,\downarrow\rangle$ and $|\downarrow,\uparrow\rangle$ states to the (0,2) singlet lifting the spin-blockade.

Results of Fig. \ref{fig7}(a) seem to be related to the recent experimental work (Ref. \onlinecite{frolov} Figure 2) that probed a wider range of magnetic field as compared to the previous experimental studies.\cite{nadj-perge,schroer,nadj2012,petersson,berg} Work [\onlinecite{frolov}] deals with EDSR involving dynamical nuclear polarization that compensates for the $g$-factor gradient within the structure. In such a case the two lines at the diagonal of the plot merge into a single resonance line as presented in Fig. \ref{fig7}(c) due to degeneracy of $|\downarrow,\uparrow\rangle$, $|\uparrow,\downarrow\rangle$ states. Although our modeling neglects the hyperfine field, our results indicate that the prominent features of the experimental data -- in the background of EDSR spectra -- are related to the ground-state singlet-triplet transition [the appearance of line that corresponds to ($\square$) direct transition] that in the present results occurs near $B=0.1$ T. Note, that the critical B for the singlet-triplet transition in our modeling is lower due to higher value of the $g$-factor in InSb.

\section{Conclusions}
We have investigated the role that spin relaxation and EDSR play in lifting the spin blockade of the current flowing across double nanowire quantum dots in the presence of spin-orbit itneraction. We found that spin relaxation mediated by phonons leads to a spontaneous lifting of the spin blockade. In consequence the resonant lifting of the Pauli blockade is observed only for a single triplet state -- the one with the spins antiparallel to the external magnetic field.
The change of the ground state in higher magnetic fields from the singlet to the $|\uparrow,\uparrow\rangle$ triplet leads to an effective spin blockade of both spin polarized triplets.
 This  leads to an appearance of additional resonant lines with AC induced transition to (0,2) singlet that do not involve phonon mediated relaxation and which are distinctly present in the recent experimental results.\cite{frolov}

\section*{Acknowledgements}
This work was supported by the funds of Ministry of Science and Higher Education (MNiSW) for 2012 -- 2014 under Project No. IP2011038671, and by PL-Grid Infrastructure. M.P.N. gratefully acknowledges the support from the Foundation for Polish Science (FNP) under START and MPD programme co-financed by the EU European Regional Development Fund.


\begin{thebibliography}{00}
\bibitem{comm} In the presence of spin-orbit coupling the singlet and triplet character of the states is preserved only approximately.
\end{thebibliography}
\end{document}